# Lunatic Stocks:
# Moon Phases as Irregular Sampling Features
# for Pattern Recognition in the Stock Markets

Luis A. Mateos

*Abstract*

This paper presents a novel idea on incorporating the Moon phases to the classic Gregorian (Solar) calendar time sampling methods for finding meaningful patterns in the stock markets.

The four main Moon phases (New Moon, First quarter, Full Moon and Third quarter) are irregular in time but with well defined sampling structure as the Moon orbits the Earth completing its period. A Full Moon may appear in one month of the year on the 2nd, on the next month the Full moon may appear on the 4th and in the next ten years on the 13th of the same month. This structure which is irregular in time makes it interesting to study together with the stock market data. Moreover, the moon affects multiple physical things on the earth, such as the ocean tides, the behavior of living organisms as well as humans mood and decision when risking and investing.

## I. INTRODUCTION

The golden grail for computer predictive systems in the stock markets is to predict accurately when to buy and when to sell assets, regardless of the unexpected ups and downs on the markets.

However, predicting the stock markets has been hard and perhaps it is an impossible task. There are so many elements and circumstances involved in the outcome of stock prices, such as: global / local political, the pandemic (COVID-19 / Delta / Omicron) economical and environmental factors to mention a few. In addition, the individual investors add an extra factor to the common trends in the markets. These individual investors may base their decisions on news, fake news, the place they worked, gossip from friend, the mood they were on that day; without making any rigorous analysis of the markets.

Even if some financial institutions use strategic mathematical methods which incorporate most of the factors affecting the stock markets and can predict the trend of the markets in some way. The individual investor behavior can affect the predicted trend unexpectedly. Nowadays, anybody with a cellphone and a *market app* can "play" (invest) in the stock markets and alter the market unexpectedly. One recent example was the exponential increase in the Game Stop - GME stock in early 2021, followed by AMC hike, [2][3].

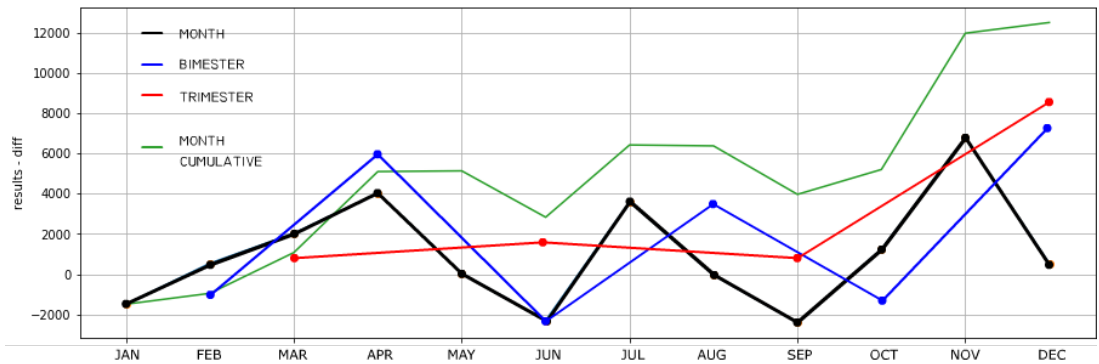

Fig. 1. Historically best and worse months in the major USA stock markets (data from DOW index 2000 - 2021).

Thus, there will not be an accurate predictive system (with or without AI) for stock markets until it can understand all the variables and factors affecting each of the investors and their reason for buying and selling stocks. Nevertheless, even in random data it is possible to extract patterns and the same apply to the stock markets. It is well know that historically the best months in the stock market are April and November, as these months have been registering the peaks in the markets and that June and September have been historically the worst months in the main stock markets, see Fig. 1.

Email: lamateos@mit.edu

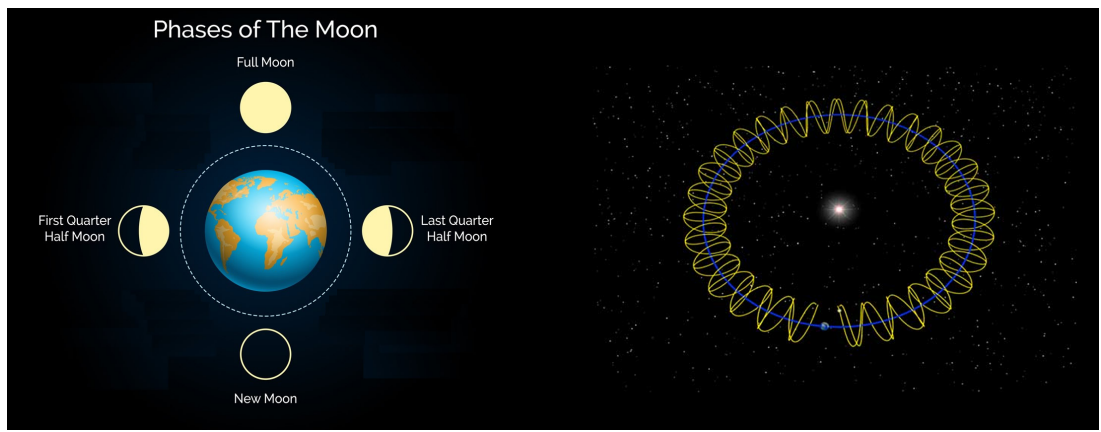

Fig. 2. Main phases of the Moon (right). Irregular periods from the Moon phases (left).

### A. Irregular Time Series Analysis

It is common to analyze stock markets data with time series. The data can be plotted against well defined sampling variables based in the Gregorian calendar (Solar calendar) to understand the markets changes in a daily, weekly, monthly, yearly basis and any other combined or derived time metrics such as number of week, weekday, number of day in year, season (spring, summer, fall, winter), etc.

However, if we incorporate a well define structure but with irregular time periods such as the Moon phases, we are not only adding a different time series. We are adding a time series with multiple properties which affect the humans physical and physiological [3]. In this sense, the lunar phases enrich the data and creates another layer of possibilities for finding trends in the stock markets.

### B. Moon Phases

There are four major "lunar" or Moon phases: New Moon, First quarter, Full Moon and Third or Last quarter.

*1) New Moon.:* This is the invisible phase of the Moon, with the illuminated side of the Moon facing the Sun and the night side facing Earth. In the new Moon or full Moon, the tide's range is at its maximum.

*2) First quarter.:* The Moon is now a quarter of the way through its monthly journey and you see half of its illuminated side.

*3) Full moon.:* This is as close as we come to seeing the Sun's illumination of the entire day side of the Moon. The Moon is opposite the Sun, as viewed from Earth, revealing the Moon's dayside.

*4) Third (Last) quarter.:* The Moon looks like it's half illuminated from the perspective of Earth, but really you're seeing half of the half of the Moon that's illuminated by the Sun.

These phases are not consistent in time with respect to the Solar calendar. Since, one phase on a given month may consists of 7 days and in the next month of 8 days. In the same way, a given Moon phase will appear in different day of the month. For example, the full Moon for the month of May in 2021 was the Wednesday 26th of May, whereas for the next month was the Thursday 24th of June, see Fig. 2.

The Moon affects the earth in a physical way. One example is the tide, which is at it's maximum in the new or full Moon [1]. In the same way, the Moon affects human, either from the light at night or from the gravitational pull making us sleep more or less minutes and affect the behavior [4][5].

Here I just want to comment a few situations in which the Moon affect humans as introduction. In the next section we focus on the numbers in the stock market mapped to the lunar phases together with common time sampling variables.

### C. Term Lunatic

In Shakespeare's "Othello," the maid Emilia tells Othello that the moon has drawn too close to the Earth — and driven men insane.

As far back as 400 B.C., physicians and philosophers blamed behavioral changes on the pull of the moon. The word "lunatic," after all, came from the idea that changes in mental state were related to lunar cycles.

The connection between the two is even supported in historic legal treatises: Famed British jurist William Blackstone wrote that people gained and lost their ability to reason according to the moon's shifting phases.

The possibility that humans could be affected by the moon's cycles is not entirely groundless [5].

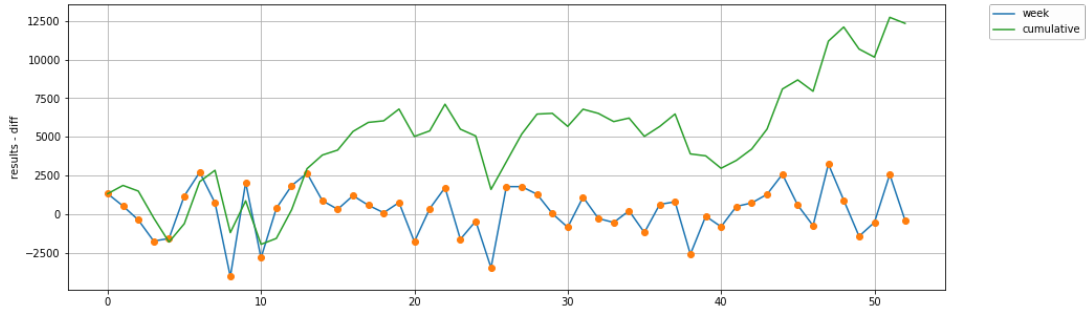

Fig. 3. Weekly sampling for the DOW index during 2000 - 2021.

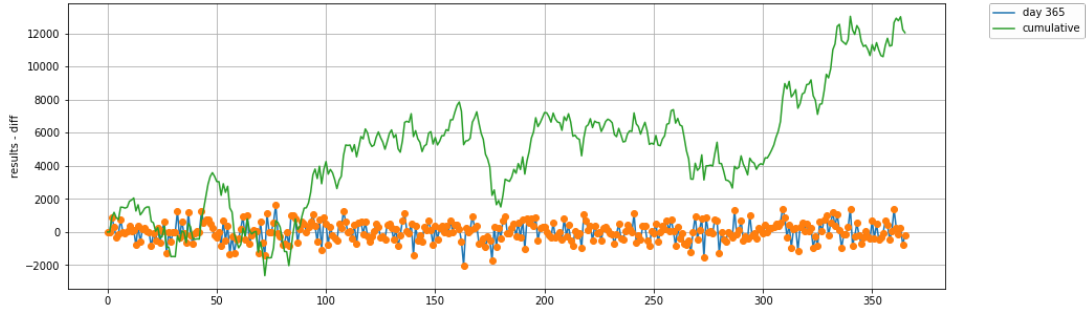

Fig. 4. Daily sampling for the DOW index during 2000 - 2021.

## II. STOCK MARKET PATTERNS

Fig. 1 shows plots from [Semester, Trimester, Bimester, Monthly] sampling periods of the DOW index during the years 2000-2021. It is possible to notice that the second half of the year is better than the first, in corroboration of this data, the fourth trimester is the best in the stock markets. Following this top-down approach reveals that the best bimester is November-December and the best month is November for the DOW stock market.

Moving to a more granular stage, in Fig. 3 the best week of the year is the 47th week, which falls in November. For the final level of sampling presented in this study the best days of the year in the stock market appear in different times of the year as shown in Fig. 4, at the beginning and at the end of the years. The day number 77 with nearly 1700 points, then day number 309 with nearly 1500 points and day number 340 with nearly 1500 points too.

If the sampling is set by day of the month, then it is possible to notice in Fig. 5 that the best day for the stock market has been the 16th. While, for investing the 19th and 20th with the lowest points.

Another interesting sampling is by the week day as shown in Fig. 6. The plot shows that Tuesdays have been the best day for the stock markets. While, Fridays the worst.

## III. TIMELY IRREGULAR WELL DEFINED PATTERNS

The data used in this study can be find in [6]. The data consist of the main Moon phases together with the dates when appearing. In the code it is shown the year, month, day and Moon phase. The Moon phases are set to numeric indexes, where 0 is New Moon, 10 is First quarter, 20 is Full moon and 30 is the Third quarter. The idea behind this numbering is that each Moon phase period is less than 10 days. Thus, it is easier to work with a single array when analyzing the data.

```
year,month,day,luna
1989,1,7,0 (New Moon)
1989,1,14,10 (First)
1989,1,21,20 (Full Moon)
1989,1,29,30 (Third)
1989,2,5,0 (New Moon)
.
```

The lunar data is filled in between the Moon phases. Resulting in the number of days that the Moon period last with its corresponding date as shown in the list below. Moreover, in this preliminary list is possible to notice the irregularity in the periods. The New Moon period last 7 days plus the New Moon, while the First quarter and Full Moon periods last only 6 days.

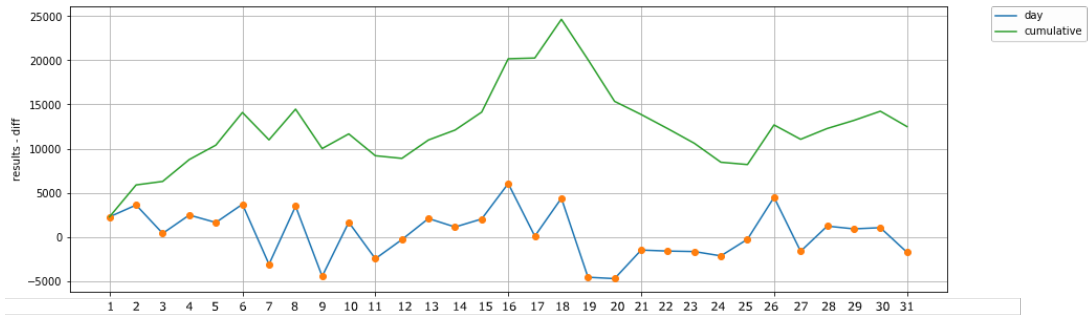

Fig. 5. Day of the month sampling for the DOW index during 2000 - 2021.

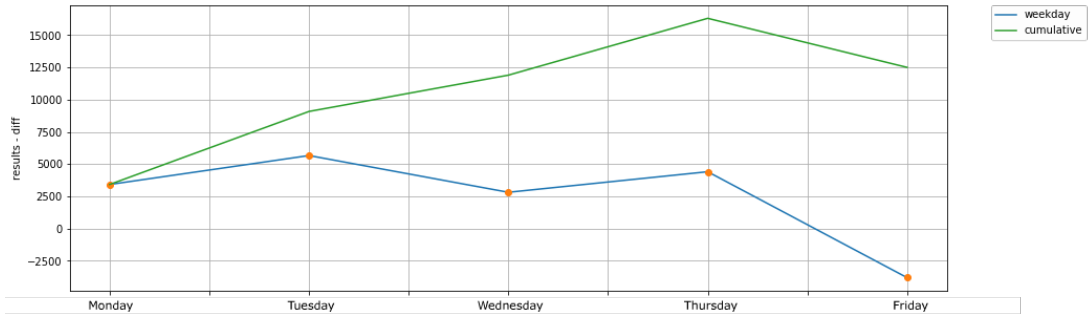

Fig. 6. Weekday sampling for the DOW index during 2000 - 2021.

```
year , month , day , luna
1992 ,1 ,4 ,0  (New Moon)
1992 ,1 ,5 ,1
1992 ,1 ,6 ,2
1992 ,1 ,7 ,3
1992 ,1 ,8 ,4
1992 ,1 ,9 ,5
1992 ,1 ,10 ,6
1992 ,1 ,11 ,7
1992 ,1 ,12 ,10  (First)
1992 ,1 ,13 ,11
1992 ,1 ,14 ,12
1992 ,1 ,15 ,13
1992 ,1 ,16 ,14
1992 ,1 ,17 ,15
1992 ,1 ,18 ,16
1992 ,1 ,19 ,20
```

```
year , month , day , luna
1992 ,1 ,20 ,21  (Full Moon)
1992 ,1 ,21 ,22
1992 ,1 ,22 ,23
1992 ,1 ,23 ,24
1992 ,1 ,24 ,25
1992 ,1 ,25 ,26
1992 ,1 ,26 ,30  (Third)
1992 ,1 ,27 ,31
1992 ,1 ,28 ,32
1992 ,1 ,29 ,33
1992 ,1 ,30 ,34
1992 ,1 ,31 ,35
1992 ,2 ,1 ,36
1992 ,2 ,2 ,37
1992 ,2 ,3 ,0  (New Moon)
.
```

The last step for preparing the entire data is to integrate the stock market data: volume, open, high, low, close, difference and derived variables. Please refer to [6].

## IV. LUNATIC STOCKS

Section II shows the different sampling periods with the Solar calendar for analyzing the stock markets during a year and related sampling such as day of the month and day of the week.

Another way to sample the stock market is by the Moon phases. This sampling is not regular as the Moon's period are inconsistent if matching to the solar calendar. Thus, by finding the best and worst sampling points of the Moon phases to the stock market data, we have another metric for obtaining insights in the markets data.

Fig. 7 shows the four Moon phases and their periods, revealing that the phase New Moon and First quarter are positive, while Full Moon and Third quarter are negative. If grouping the days per each period, the best period is the one with the Full Moon, then Third quarter, First quarter and the worst with little earning is the New Moon, see Fig. 8.

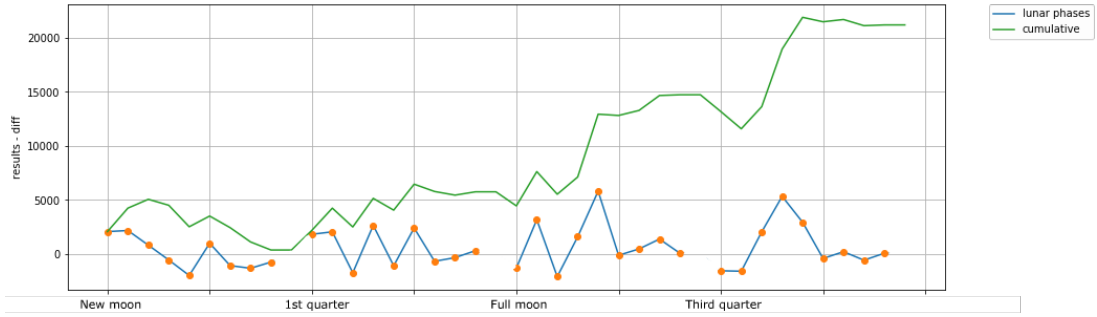

Fig. 7. Moon phases in the DOW index from 1992 to 2021.

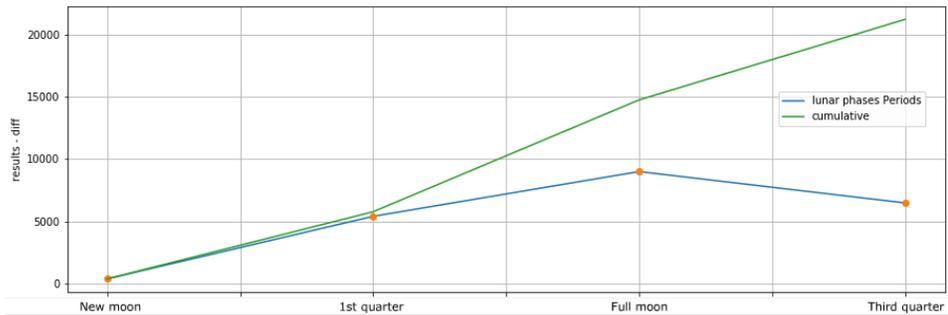

Fig. 8. DOW stock market since 1992 to 2021 by Moon phase.

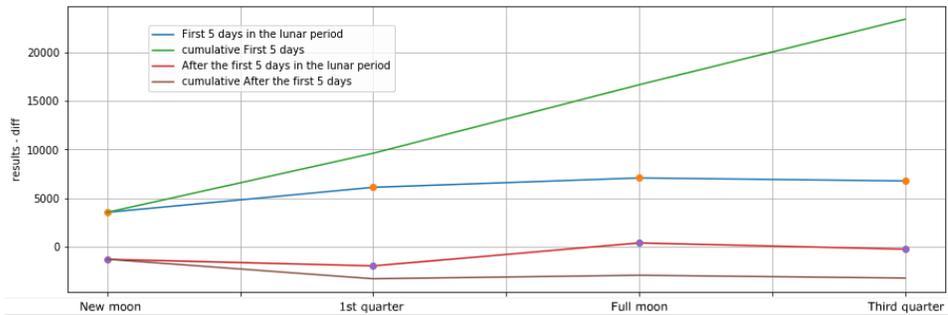

Fig. 9. First five days from each Moon period and last days from each Moon period (after the 5th day of the period) in the DOW index from 1992 to 2021.

One interesting insight in the data is that in the full Moon the markets go down, this is to contrast the common belief that full Moon is a good date for investing. Nevertheless, the days after the full Moon are the ones with the most earnings. That is why if comprising the full period of full Moon until the Third quarter gives the best earnings, see Fig. 7.

Another finding from a simple analysis to group the number of days after each Moon phase started leads to an interesting insight. If taking only the first 5 days from each period reveals a local best earnings (highest points). Whereas, if taking the days after the fifth day from each period leads to near zero or negative points in all the Moon periods see Fig. 9.

## V. CONCLUSION AND FUTURE WORK

In this paper, a novel concept of incorporating the Moon phases as irregular sampling data metric is presented. The lunar data is fused to the DOW index for finding insights in the stock markets data from a different sampling perspective to understand human behavior when investing and risking.

Lunar data is irregular in time if compared to the Solar samplings, such as day, week, month, day of the month, day of the year. A Moon period can last 7 or 8 days and can appear in different day of the month. Nevertheless, the structure of the Moon phases are wells define as the Moon is orbiting the Earth in the same pace.

In this study, the Moon phases were used as sampling for the DOW index, revealing interesting insights in the four major Moon phases: New Moon, First quarter, Full Moon and Third quarter. Moreover, if grouping the first five days from each Moon phase leads to a local maxima best earnings, whereas if taking into account the last days of each period leads to near zero points.

Besides the study of the lunar data with the stock markets. The key takeaway of this study is to add the Moon phases sampling periods to other well-known sampling period and mix-match them to obtain a better hybrid model in challenging data sets involving human or animal behaviors.